\documentclass[journal,twocolumn]{IEEEtran}

\usepackage{amsfonts}
\usepackage{times}
\usepackage{graphicx}
\usepackage{latexsym}
\usepackage{dsfont}
\usepackage{amssymb}
\usepackage{amsmath}
\usepackage{cite}
\usepackage{verbatim}
\usepackage{subfigure}

\newcommand{\figref}[1]{{Fig.}~\ref{#1}}

\def\bb0{{\mathbb{0}}}

\def\ba{{\mathbf{a}}}
\def\bb{{\mathbf{b}}}

\def\bw{{\mathbf{w}}}

\def\by{{\mathbf{y}}}

\def\b0{{\mathbf{0}}}

\def\sf0{{\mathsf{0}}}

\usepackage{epstopdf}
\usepackage{enumerate}
\usepackage{algorithmicx}
\usepackage{algorithm}
\usepackage{amsmath}
\usepackage[noend]{algpseudocode}
\usepackage{float}
\usepackage{hyperref}
\usepackage{color}
\usepackage{makeidx}
\usepackage{bbm}
\usepackage{graphicx}

\begin{document}
\title{Leveraging mmWave Imaging and Communications for Simultaneous Localization and Mapping }
\author{Mohammed Aladsani, Ahmed Alkhateeb, and Georgios C. Trichopoulos\\  Arizona State University, Email: maladsan, alkhateeb, gtrichop@asu.edu}
\maketitle

\begin{abstract}
In this work, we propose a novel approach for high accuracy user localization by merging tools from both millimeter wave (mmWave) imaging and communications. The key idea of the proposed solution is to leverage mmWave imaging to construct a high-resolution 3D image of the line-of-sight (LOS) and non-line-of-sight (NLOS) objects in the environment at one antenna array. Then, uplink pilot signaling with the user is used to estimate the angle-of-arrival and time-of-arrival of the dominant channel paths. By projecting the AoA and ToA information on the 3D mmWave images of the environment, the proposed solution can locate the user with a sub-centimeter accuracy. This approach has several gains. First, it allows accurate simultaneous localization and mapping (SLAM) from a single standpoint, i.e., using only one antenna array. Second, it does not require any prior knowledge of the surrounding environment. Third, it can locate NLOS users, even if their signals experience more than one reflection and without requiring an antenna array at the user. The approach is evaluated using a hardware setup and its ability to provide sub-centimeter localization accuracy is shown.
\end{abstract}


\section{Introduction} \label{sec:Intro}

Simultaneous localization and mapping (SLAM) is defined as the ability of a robot or a system to identify its environment, create a three-dimensional map, and acquire its
current position \cite{Durrant-Whyte2006a}. SLAM enables autonomous agents to navigate independently in a crowded environment avoiding collisions with the surroundings and protecting humans from possible injury. Additionally, SLAM has found applications in virtual and augmented reality, autonomous driving, and assisted living technologies \cite{Witrisal2016}. A common requirement for most SLAM applications is providing the location of the user/device with high spatial accuracy. Spatial resolution is directly proportional to the wavelengths of the utilized EM signals, therefore mmWave has become a promising spectrum for SLAM systems. Due to their short wavelength, millimeter wave signals promise a high spatial resolution \cite{Trichopoulos2013,Sheen2001}.

\begin{figure}[t]
	\vspace{15pt}
	\centerline{
		\includegraphics[width=1\columnwidth]{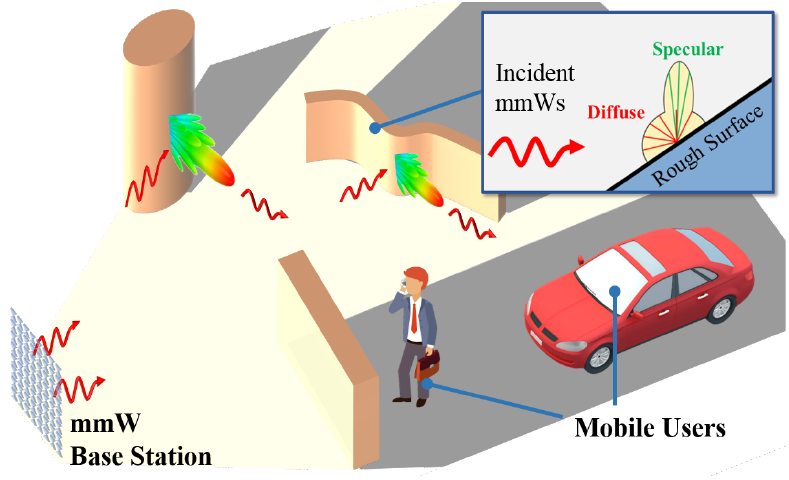}
	}
	\caption{Synergy of mmWave imaging and communications for SLAM. Exploiting the scattering properties at mmWave, both LOS and NLOS objects can be accurately imaged and localized.}
	\label{fig:Intro_fig}
\end{figure}

One of the early work on SLAM in wireless network environment was done by using the WiFi Received Signal Strength Indicator (RSSI) from multiple access points (AP) and estimating the location of a mobile station (MS) \cite{Dumont2014}. Othersextended SLAM into mmWave networks by incorporating direction of arrival (DoA) and creating a fingerprinting database for possible locations \cite{Wei2017}. Sub-6 GHz systems have been used co-currently with mmWave systems by estimating Angle of Arrival (AoA) from the sub-6 GHz system and feeding it to the mmWave system to simplify the beam
training for ranging and localization \cite{Maletic2018}. Algorithms leveraging Angle Difference of Arrival (ADoA) were developed by first estimating the position of access points
then estimate the location of the user by comparing the ADoA from each AP to the user. However, those methods require at least 4 APs \cite{Palacios2018,Palacios2017}. At mmWave frequencies, localization algorithms for LOS and NLOS were developed in \cite{Guidi2016,Shahmansoori2018}. These solutions, however, are limited for only one reflection and require an antenna array at the mobile user.  

Thanks to the hundreds to thousands of antennas that are expected to be deployed at mmWave base stations (BS) \cite{Rebeiz2015,MIDAS} ,  these BSs could also be used for mmWave imaging. Further, future communication systems “6G and beyond” will likely utilize even higher frequency bands \cite{Xing2018}. The higher bandwidth can be leveraged to provide even higher resolution images thus increasing the accuracy. In this paper, we develop a novel localization and mapping solution by leveraging tools from both mmWave imaging and communications. \textbf{Using mmWave imaging, we can enable simultaneous localization and mapping without a priori knowledge of the geometry and material properties of the surroundings. Further, the proposed solution requires only a single BS with a single antenna array and does not require the user to have more than one antenna. Also, it can account for multiple reflections in a single travel path, which is the first work, to the best of the authors’ knowledge, that proposes a solution to this problem.}

\begin{figure}[t]
	\centerline{
		\includegraphics[width=1\columnwidth]{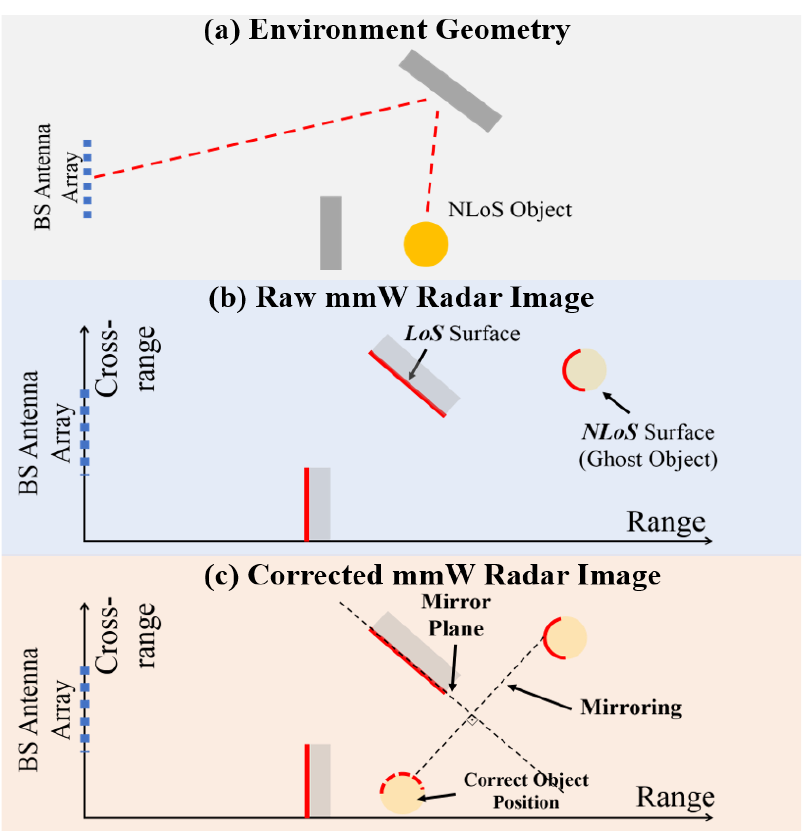}
	}
	\caption{Image correction for NLOS objects. (a) A simple topology with two
		LOS and one NLOS objects. (b) In the mmWave radar image, the NLOS appears behind the LOS surface. (c) Using mirror translation, the NLOS
		is placed in the correct position.}
	\label{fig:NLOS_correction}
\end{figure}

\section{NLOS AND LOS IMAGING USING MMWAVE}

The process of the proposed SLAM method involves three steps: (i) mmWave imaging of the surrounding geometry including both LOS and NLOS areas, (ii) estimating the user AoA and ToA using a wideband uplink pilot signal, and (iii) fusion of the mmWave map, AoA, and ToA to determine the location of the user. One of the key enablers in this approach is the ability of the system to provide images for NLOS areas.

\subsection{Holographic mmWave Imaging}

In this work, the mmWave images are obtained using a monostatic synthetic aperture radar (SAR) imaging method \cite{Sheen2001}, \cite{Soumekh1999}. We assume a planar uniform array of isotropic
antennas that are connected to ideal transceivers operating on a wide frequency bandwidth. Using holographic imaging the obtained 3D image is \cite{Sheen2001}:
\begin{equation}
f(x,y,z)=\text{IFT}_\mathrm{3D} \left\{\text{FT}_\mathrm{2D} \left\{ R(x,y,f)\right\}  e^{-j\sqrt{4k^2-k_x^2-k_y^2} Z} \right\}   
\end{equation}

Where $f(x,y,z)$ is the image of the environment in cartesian coordinates, FT and IFT denote the discrete spatial Fourier and inverse Fourier transforms. Further,  $R(x,y,f)$ is the scattered signal over a given range of frequencies $f$  collected at each $(x, y)$ SAR position, $k$ is the wave number and $k_x$ and $k_y$ represent the spatial frequency of $x$ and $y$ . 

This simple SAR reconstruction approach could also be implemented using the antenna array of a BS. However, other imaging reconstruction methods (e.g. beam scanning, multistatic imaging) can also be adopted depending on the BS topology (e.g. analog, digital, or hybrid phased array). In any case, the properties of the mmWave 3D map of the surroundings will be limited by the (i)operating frequency, (ii) bandwidth, and (iii) antenna array aperture of the BS.

\subsection{Image correction using mirroring for NLOS objects}

The 3D images acquired using the radar imaging do not represent the actual geometry of the surroundings due to the multiple bounces of mmWaves. As such, the radar images contain information for both LOS and NLOS objects and proper image correction is needed. Our group has already proposed a correction algorithm by using mirroring techniques \cite{Doddalla2018}. The algorithm is based on the assumption that objects are opaque at mmWaves, therefore any object that appears behind a surface on the radar images is considered an artifact and needs to be corrected (see \figref{fig:NLOS_correction} (b)). The first step of the algorithm is to identify objects behind LOS objects and mark them for correction. As such, every LOS surface is assumed a mirror surface and every obstructed pixel is mirrored around that surface, as illustrated in \figref{fig:NLOS_correction} (c). If the LOS surface is curved, then it is discretized into planar segments and the process is repeated accordingly for every segment. Diffraction at the edges of the surfaces is assumed to be on the same level as diffuse scattering and is not treated separately.


\begin{figure}[t]
	\centerline{
		\includegraphics[width=1\columnwidth]{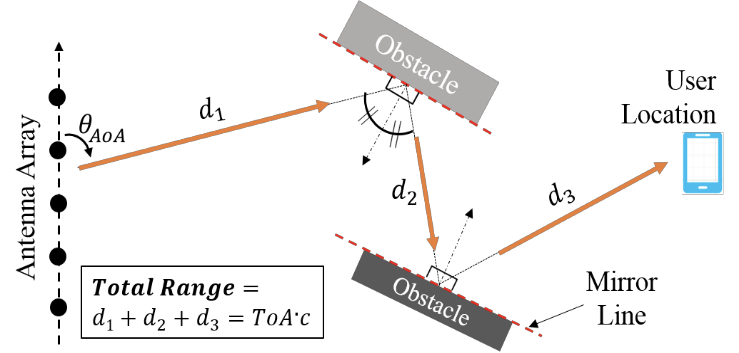}
	}
	\caption{Ray casting to locate the user in the mmWave image. The length of
		the ray path is equal to ToA multiplied by the speed of light, $c$.}
	\label{fig:Raytracing}
\end{figure}

\section{Finding the User Location}
To locate the user, we first estimate the AoA and ToA of the dominant paths of the channel between the BS and the user. Then, leveraging the knowledge of the environment’s geometry, we project the estimated AoA and ToA on the acquired mmWave image to localize the user.

\subsection{Estimating AoA and ToA}
Consider a uniform linear antenna array (ULA) with N elements
placed along the cross-range axis, and with uniform $\lambda/2$ spacing. For simplicity of exposition, and without LOSs of generality, we assume that only one path exists for the channel between the user and the antenna array. This path can be eith a LOS or NLOS path. If the user transmitted an uplink pilot signal $s_p\left(f\right)$ at frequency $f$, then the received signal at the antenna array can be expressed as
\begin{equation}
\by(f)=\alpha_u  s_p\left(f\right) \ba\left(\theta_u\right)
\end{equation}
where $\alpha_u$ and $\theta_u$ are the complex gain and AoA of the channel path between the array and the user. Further, the array response vector of this AoA is defined as $\ba\left(\theta_u\right)=\left[1, e^{j kd \cos(\theta_u)}, ..., e^{j kd (N-1) \cos(\theta_u)}\right]^T$. At the receive array, a joint search over the angles of arrival and ranges is performed to estimate the user AoA and ToA (or equivalently the range). For this, the received signal $\by(f)$ is combined by a steering vector $\bw(\theta)$ with sweeping over the steering angle $\theta$. The combined received signal at frequency $f$ and steering angle $\theta$ can then be written as
\begin{equation}
y_c\left(\theta,f\right) = \bw(\theta)^T \by(f).
\end{equation}
To transform this combined received signal to the range domain, we apply the inverse spatial Fourier transform \cite{Sheen2001}
\begin{equation}
y_c\left(\theta,r\right) = \mathrm{IFT} \left\{y_c\left(\theta,f\right) e^{-jkr}\right\}
\end{equation}
By scanning this received combined signal $y_c\left(\theta,r\right)$ over all the angle/range points, we estimate the user AoA,$\hat{\theta}_u$, and range, $\hat{r}_u$ (or equivalently the ToA, $\hat{T}_u$).  Note that when the user is in NLOS, the range $\hat{r}_u$ does not represent the actual distance of the user from the array center, but rather the total travelling distance from the user to the array through the (multple) reflections. 

\subsection{Projecting AoA and ToA on the 3D mmWave images}

The 3D mmWave image will be combined with the estimated AoA and ToA to find the location of the user with respect to the environment. The physical center of the BS antenna array is used as a reference point for the AoA and ToA. Namely, a ray is launched from the BS center at an angle equal to the estimated AoA $\hat{\theta}_u$. For LOS users, the length of the ray will be equal to $\hat{r}_u = \hat{T}_u  c$, where $c$ is the speed of light and the user will be located at the end of the ray path. If the user is in NLOS, then the ray will intersect a mirror object at a distance less than $\hat{r}_u$. Then, a second ray is launched from the intersected point at the specular angle with respect to the mirror surface. This is repeated for multiple bounces until the total length of all the rays is equal to $\hat{r}_u$. The end of the ray path is then the location of the user, as illustrated in \figref{fig:Raytracing}. The accuracy of the proposed algorithm will depend on the fidelity of the mmWave images and accuracy of the estimated AoA/ToA. Both steps depend on the (i) operating frequency, (ii) bandwidth,  and (iii) BS antenna array aperture. 

\begin{figure}[t]
	\centerline{
		\includegraphics[width=.9\columnwidth]{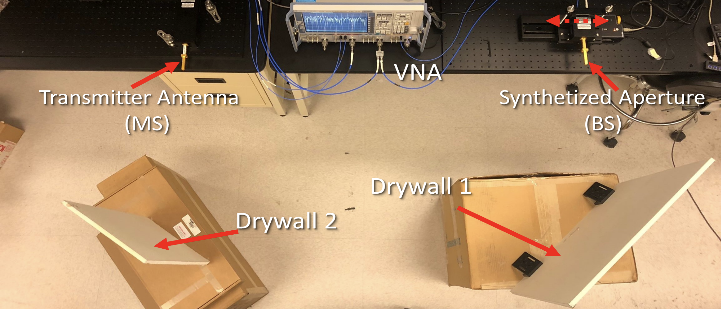}
	}
	\caption{Experimental setup. Drywall 1 is in LOS and drywall 2 in NLOS.
		The user is placed in NLOS besides drywall 2.}
	\label{fig:Fig4}
\end{figure}

\section{mmWave SLAM Experiment for A 2D Environment }

As a proof of concept, the proposed SLAM method is evaluated using a 2D mock-up environment in the 220-300 GHz range. Due to the lack of a full-fledged phased array system at mmWave frequencies, a synthetic aperture radar (SAR) is implemented to emulate the imaging performance of a BS with a 13-cm aperture linear antenna. As such, the SLAM problem is limited to two dimensions (range and crossrange). The mock-up environment is comprised of two pieces of drywall one is in LOS and the other is NLOS, as depicted in \figref{fig:Fig4}.

\begin{figure}[t]
	\centerline{
		\includegraphics[width=1\columnwidth]{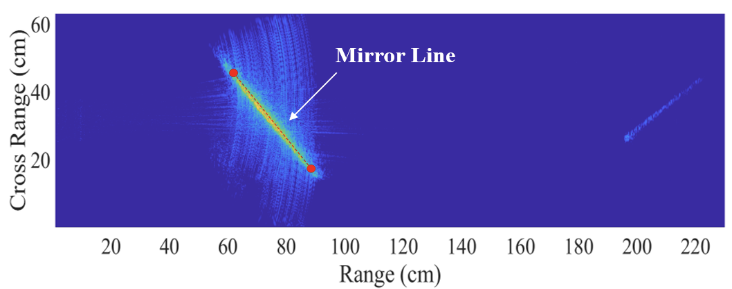}
	}
	\caption{A 2D image of the environment. The second drywall is in
		NLOS and wrongfully appears behind the first drywall.}
	\label{fig:Fig5}
\end{figure}

\begin{figure}[t]
	\centerline{
		\includegraphics[width=1\columnwidth]{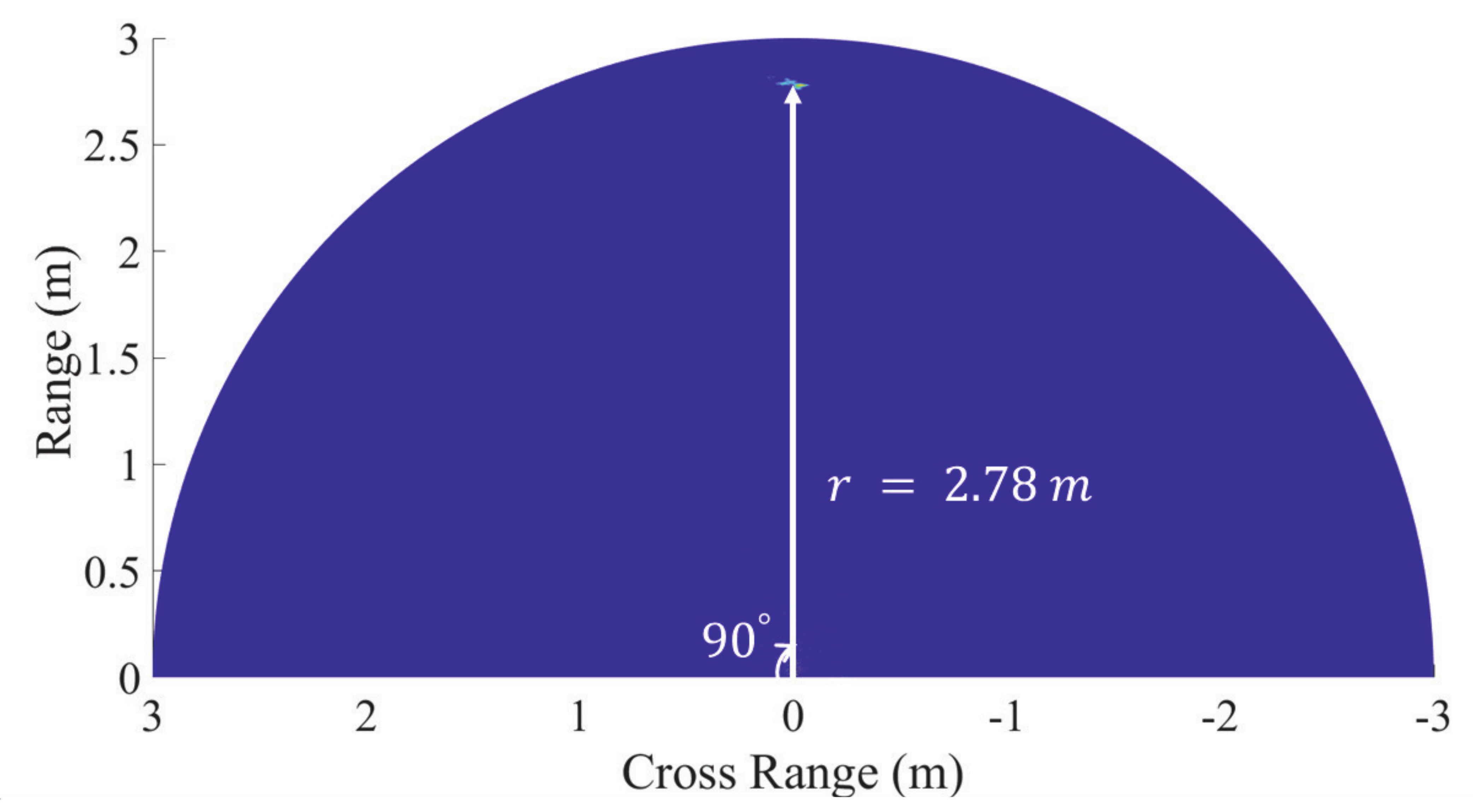}
	}
	\caption{Map of range and cross-range based on the AoA and ToA
information. Due to the NLOS location, the user appears wrongfully  in boresight with respect to the receiving aperture.}
	\label{fig:Fig6}
\end{figure}

For SAR, two Vector Network Analyzer (VNA) extenders \cite{VD} coupled to diagonal horn antennas are used to form a monostatic radar configuration with the support of a motorized translation stage to scan the aperture. The computer-controlled SAR systems record the S21 at multiple frequency points in the 220-300 GHz range. Using holographic image reconstruction [19] and image correction, a 3D map of the environment is acquired, as shown in \figref{fig:Fig5}. It is noticed that the geometry of both LOS and NLOS objects are in excellent agreement with the actual topology. The smearing that appears around the first wall is due to the phase error introduced by the cables as the VNA extender moves on the translation stage. Subsequently, to emulate the mobile user, a pilot signal is transmitted by a VNA extender transmitter (Tx) when placed beside the NLOS drywall. The extender is oriented such that the antenna is facing the drywall at approximately $45^\circ$. The Tx module transmits multiple tones in the 220-300 GHz range and the scattered signals are received by the SAR receiver module ($S_{21}$). Then, using standard  beamforming techniques the AoA and ToA are estimated from the recorded $S_{21}$ signals. The map of the AoA and ToA is plotted in \figref{fig:Fig6}. Due to the multiple scattering of the mmWave signals on the drywall surfaces, the user appears in a single location, but no information on the surrounding geometry is provided in this image. Finally, every point on the AoA-ToA diagram is projected on the 2D map, as described in Section 3.

\begin{figure}[t]
	\centerline{
		\includegraphics[width=1\columnwidth]{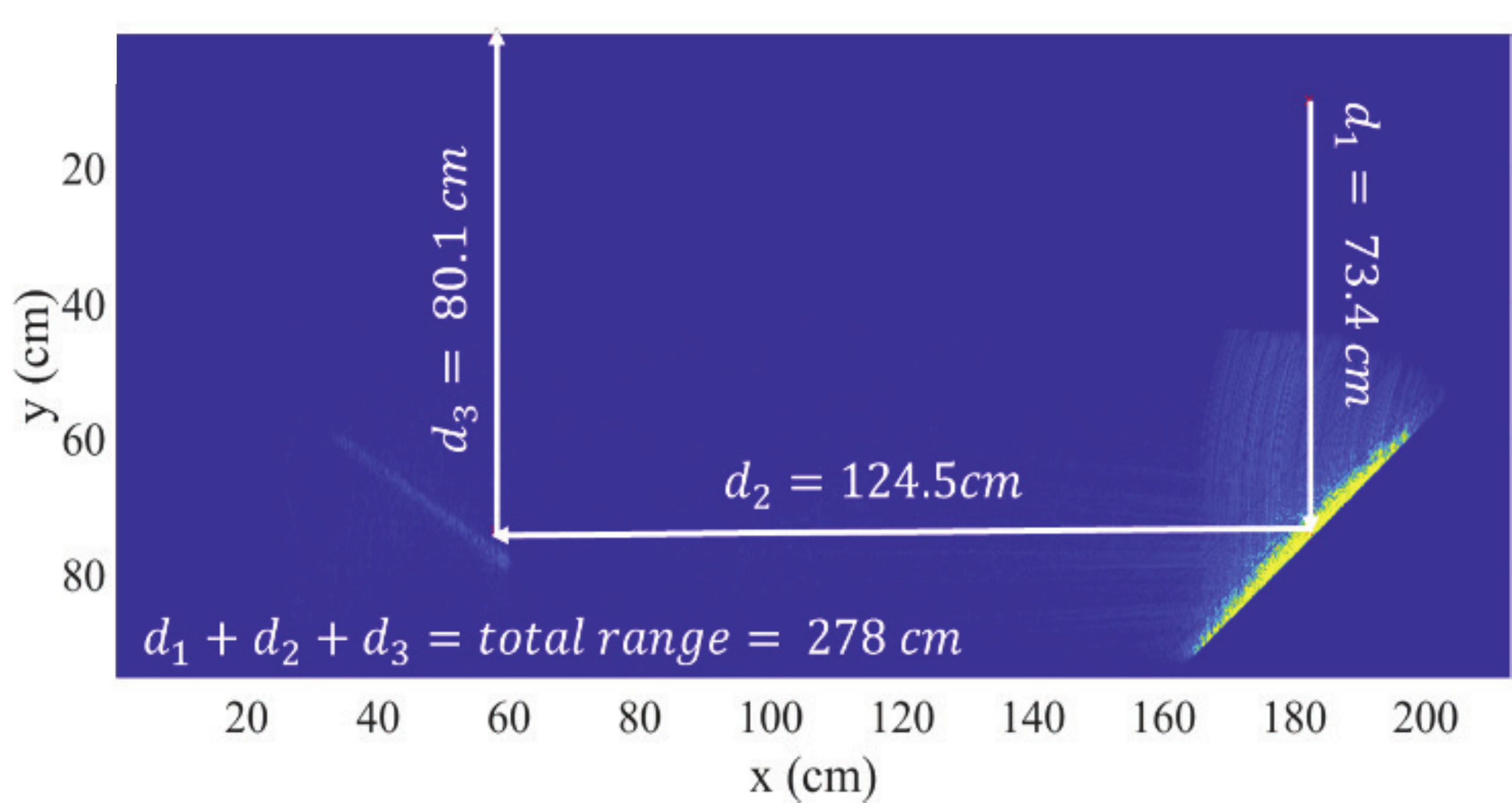}
	}
	\caption{Projection of the user location on the corrected 2D mmWave image}
	\label{fig:Fig7}
\end{figure}

\figref{fig:Fig7} shows the location of the user inside in the reconstructed 2D map. The method exhibits sub-centimeter accuracy in estimating the location of the user on the y-axis. However, the accuracy on the x-axis depends on the estimation of the mirrors’ orientation. In this work, the orientation is manually defined by choosing two points at the ends of each drywall to determine the mirror lines. Due to the high spatial resolution of the 2D image, the mirror lines match the actual drywall orientation. In this ideal situation, the positioning accuracy in the x-axis is only limited by the cross-range resolution of the system, which at a range of 2.8m corresponds to the minimum resolvable length of 2.6 cm. In practical scenarios, several parameters can differ from the laboratory conditions demonstrated in this first proof of concept of SLAM using NLOS mmWave imaging methods.

\section{Discussion}
Mobile users can also be equipped with an omnidirectional antenna, which will increase the possibility of having a multipath reflection, thus multiple AoAs. In such scenario, mapping into the mmWave image needs to be repeated for all AoAs. Additionally, in this work, ToA is estimated using stepped frequency of a significantly wide bandwidth (~100 GHz). As such, ToA estimation is expected to be limited by the bandwidth limitations of the wireless channel and hardware. Furthermore, for more accurate estimation of AoA, more sophisticated methods, compared to the one adopted in this proof of concept, can be used. Finally, mirror plane estimation is crucial for accurate user localization and further investigation is needed to implement automated methods for surface geometry reconstruction.

\section{Conclusion}
In this paper, we proposed the synergy of mmWave imaging and  communications for highly accurate SLAM. Leveraging the geometry side information captured by mmWave imaging, we demonstrated SLAM for NLOS users with multiple reflections. A simple 2D scenario was investigated in the 220-300 GHz range using a synthesized 13-cm linear antenna array. The method successfully resolves the geometry of the surroundings and localizes the user with subcentimeter accuracy.


\end{document}